\documentclass[letterpaper,twocolumn,10pt]{article}

\usepackage{usenix}

\usepackage{float}
\usepackage{amsmath}
\usepackage{xspace}
\usepackage{siunitx}
\usepackage{enumitem}
\usepackage{listings}
\usepackage{tikz}
\usetikzlibrary{shapes,arrows,positioning,fit}
\usepackage{subcaption}
\usepackage{tabularx}
\usepackage{balance}
\usepackage{url}
\usepackage{comment}

\usepackage{hyperref}
\usepackage{xurl} 
\usepackage{cleveref}

\usepackage{booktabs}

\newcommand{\sys}{gpu\_ext\xspace}
\newcommand{\sysdriver}{\sys-Driver\xspace}

\crefname{algocf}{algorithm}{algorithms}
\Crefname{algocf}{Algorithm}{Algorithms}
\crefformat{section}{\S#2#1#3}
\Crefformat{section}{Section~#2#1#3}
\crefformat{subsection}{\S#2#1#3}
\Crefformat{subsection}{Section~#2#1#3}
\crefformat{subsubsection}{\S#2#1#3}
\Crefformat{subsubsection}{Section~#2#1#3}
\crefformat{figure}{figure~#2#1#3}
\Crefformat{figure}{Figure~#2#1#3}

  {\begin{enumerate}[itemsep=-0pt, parsep=0pt, topsep=0pt, leftmargin=2pc]}
  {\end{enumerate}}

  {\begin{list}{$\bullet$}%
    {\setlength{\parsep}{0pt}%
      \setlength{\topsep}{0pt}%
      \setlength{\leftmargin}{2pc}%
      \setlength{\itemsep}{1pt}}}
  {\end{list}}

\begin{document}

\pagestyle{plain}

% Do not print date
\date{}

% \title{\sys: An eBPF Extensible OS Policy Interface for GPU Systems}
\title{\sys: Extensible OS Policies for GPUs via eBPF}

\author{
{\rm Yusheng Zheng}$^{=,1,2}$,
{\rm Tong Yu}$^{=,1}$,
{\rm Yiwei Yang}$^{2}$,
{\rm Minghui Jiang}$^{1,3}$,
{\rm Xiangyu Gao}$^{4}$,
{\rm Jianchang Su}$^{5}$,\\
{\rm Yanpeng Hu}$^{6}$,
{\rm Wenan Mao}$^{3}$,
{\rm Wei Zhang}$^{5}$,
{\rm Dan Williams}$^{7}$,
{\rm Andi Quinn}$^{2}$\\[1ex]
$^{1}$eunomia-bpf,
$^{2}$UC Santa Cruz,
$^{3}$Alibaba Group,
$^{4}$University of Washington,\\
$^{5}$University of Connecticut,
$^{6}$ShanghaiTech University,
$^{7}$Virginia Tech
}

\maketitle
\let\thefootnote\relax\footnotetext{$^{=}$Equal contribution.}

% === Original structure ===
\begin{abstract}
Performance in modern GPU-centric systems increasingly depends on resource management policies, including memory placement, scheduling, and observability. However, uniform policies typically yield suboptimal performance across diverse workloads. Existing approaches present a tradeoff: user-space runtimes provide programmability and flexibility but lack cross-tenant visibility and fine-grained control of hardware resources; meanwhile, modifications to the OS kernel introduce significant complexity and safety risks. To address this, we argue that the GPU driver and device layer should provide an extensible OS interface for policy enforcement. While the emerging eBPF technology shows potential, directly applying existing host-side eBPF is insufficient because they lack visibility and control into critical device-side events, and directly embedding policy code into GPU kernels could compromise safety and efficiency.

We propose \sys, an eBPF-based runtime that treats the GPU driver and device as a programmable OS subsystem. \sys extends GPU drivers by exposing safe programmable hooks and introduces a device-side eBPF runtime capable of executing verified policy logic within GPU kernels, enabling coherent and transparent policies. Evaluation across realistic workloads including inference, training, and vector search demonstrates that \sys improves throughput by up to $4.8\times$ and reduces tail latency by up to $2\times$, incurring minimal overhead, without modifying or restarting applications.
% \xiangyu{I think it is better to reduce the abstract content by 50-60\% by removing some example details.}
% \yusheng{should we use general "policy" or make it resources management policy? we are also doing  optimization for a single application and doing observability/tracing}
% \JC{
% Resource management policies have become crucial in modern GPU systems, as diverse workloads contend for limited resources in multi-tenant environments. However, current approaches to policy specification and enforcement are inadequate. Policies are either hard-coded into opaque drivers or scattered across user-space runtimes, lacking flexibility, coordination, and strong safety guarantees.

% We propose exposing the GPU driver and device layer as a programmable resource management substrate using eBPF, enabling application-transparent policies with stack-wide coordination, full hardware control, and strong safety guarantees. \sys introduces a verifiable runtime that executes management policies within the GPU stack, ensuring policy code is checked for bounded execution and memory access. This allows policies to dynamically adapt to workload requirements while guaranteeing system stability. Our evaluation on real-world workloads shows that \sys improves

% }
\end{abstract}

\section{Introduction}

% 1. (General problem) GPU is important (1). Motivate the resources management policy is important (resource mpolicy has emerged as ... 
% 2. (The problem with current works) Current design rigiet for us to create new resource management policy, some one has try to modify drivers; others are trying to do it in usersapce framwork
% 3. Our key insight is that the GPU driver is uniquely positiooned in the GPU stack to have transparency into all applications, etc. 
% 4. (Our claim) In this paper, we present \sys{}, a new system that enables flexible resource management policies for GPU deployments.  Building on our key insight, \sys{} implements a stable set of control plane hooks that allow for flexible resource managment policies...
% 5. To support flexibile policies, applications must be able to observe an application's behavior on the GPU device. Safety is required... Thus, we implement our SIMT verifier. 
%6. Finally, to allow the driver and device components to communicate, we need a communication substrate.
%7. We implemented \sys{} on.... We evaluated on xyz workloads and our results are awesome. 
%8. Our contributions are as follows: 
%   a. 

% 1. (General problem) GPU is important (1). Motivate the resources management policy is important (resource mpolicy has emerged as ...
The performance and efficiency of GPU-based systems increasingly depend not only on raw compute and memory performance but also on resource management policies. GPUs account for a growing share of compute resources in modern systems~\cite{jeon2019analysis}, supporting diverse workloads such as latency-sensitive inference~\cite{kwon2023efficient,agrawal2024taming,yu2024orca}, compute-intensive training~\cite{shoeybi2019megatron,rasley2020deepspeed}, graph analytics~\cite{lin2024towards,chen2020pangolin,wang2016gunrock}, vector search~\cite{pan2024survey,cao2023gpu}, and data visualization~\cite{moreland2016vtk}, each imposing distinct and often conflicting resource requirements. GPU software stacks, spanning user-space frameworks, OS kernel drivers, vendor firmware, and device-side kernel code, necessitate coordination of memory placement and migration~\cite{sun2024llumnix,zhong2024distserve,patel2024splitwise,strati2024orion,sheng2023flexgen}, kernel scheduling in multi-tenant environments~\cite{fu2024serverlessllm,shubha2024usher,hu2023lucid,li2022miso,jayaram2023sia,xiao2018gandiva,narayanan2020heterogeneity}, and dynamic observability for online adaptation~\cite{agrawal2024vidur}. These policies must evolve alongside changing hardware architectures, emerging workloads, and varying service-level agreements (SLAs)~\cite{kwon2023efficient,yu2020fine,delimitrou2013paragon,delimitrou2014quasar,lo2015heracles,qiao2021pollux}.

Given these diverse and evolving requirements, no single policy fits all scenarios. Recent work confirms that default Unified Virtual Memory (UVM) placement policies~\cite{wang2024suv,lin2025forest,kim2025dream,liao2024svm,park2025helm} and scheduling policies~\cite{wang2024gcaps,fan2025gpreempt,kato2011timegraph,shen2025xsched} can perform poorly across workloads and lack multi-tenant coordination on different deployments.

% 2. (The problem with current works) Current design rigiet for us to create new resource management policy, some one has try to modify drivers; others are trying to do it in usersapce framwork
Unfortunately, existing programmable GPU resource management policy solutions force developers into tradeoffs between user-space methods that lack global visibility and control over system-level GPU behaviors, and kernel-level implementations that introduce complexity and safety risks. 
Concretely, user-space frameworks~\cite{ng2023paella,shen2025xsched,gim2025pie,chen2025ktransformers} offer programmability, but lack cross-application visibility and management of low-level GPU mechanisms (e.g. thread timeslices and replayable page faults~\cite{fan2025gpreempt,nvidia-open-gpu}). These approaches bind to specific runtime frameworks, requiring application-level code changes and limiting portability and reuse. Conversely, kernel-based resource management approaches~\cite{kato2011timegraph,kato2012gdev,kernelintercept,fan2025gpreempt,wang2024gcaps} provide visibility and control of GPU hardware states, their direct modification on static, vendor-specific kernel modules making them difficult to deploy and maintain, and posing stability concerns. GPU profiling frameworks~\cite{cupti,villa2019nvbit,neutrino}, including our prior workshop version~\cite{egpu}, enable kernel-level observation but lack runtime programmability and driver integration, limiting adaptive resource management.

% 3. Our key insight is that the GPU driver is uniquely positiooned in the GPU stack to have transparency into all applications, etc.
To address this, we argue that GPU resource management requires \textbf{an 
OS policy interface for GPU} providing safe, flexible and transparent programmability. The GPU driver layer is positioned to offer global visibility, enable cross-tenant coordination, and provide fine-grained management of privileged low-level GPU mechanisms without requiring application modifications. Recent CPU-side frameworks, such as eBPF, have demonstrated the feasibility of dynamic, kernel-level policy programmability with safety guarantees~\cite{bachl2021flow,Zhong22,jia2023programmable,schedext-docs,zussman2025cache_ext}. Inspired by its success, we propose treating GPU resource management as a first-class programmable OS subsystem. However, naively extending host-only eBPF remains insufficient because scheduling and memory management decisions such as thread-block dispatch, warp-level execution scheduling, and memory synchronization occur within GPU hardware and remain invisible and uncontrollable from the host. Thus, a practical solution must expose programmable policy hooks within GPU drivers and execute verified policy logic inside GPU kernels.

We present \sys{}, a cross-layer policy runtime that transforms the GPU driver and device into a programmable OS subsystem. \sys{} exposes stable, verified control-plane hooks within the GPU driver, enabling safe and customized programmable resource management policies. 
\sys{} further supports injecting verified eBPF policy logic into GPU kernels, providing consistent and transparent policy enforcement across the host-device boundary.

Designing \sys{} has three technical challenges to solve: \textbf{C1:} Exposing low-level driver mechanisms provides expressiveness but risks instability; we balance this tension with a narrow policy interface built around hardware-aligned abstractions (\S\ref{sec:design-interfaces}). \textbf{C2:} Scalar eBPF semantics conflict with GPU SIMT parallelism, causing divergence and deadlocks. Thus, we introduce a SIMT-aware verifier and warp-level execution model for optimization (\S\ref{sec:verification}). \textbf{C3:} Host-device memory hierarchies differ in latency and consistency; we bridge them with cross-layer eBPF maps using relaxed consistency (\S\ref{sec:verification}).

We implement \sys{} on Linux by extending the NVIDIA open GPU kernel modules~\cite{nvidia-open-gpu} and providing an eBPF runtime for GPU devices. We evaluate \sys across GPU workloads including LLM inference pipelines, GNN training, vector search, and mixed-priority multi-tenant scenarios. Using \sys, we implement adaptive memory prefetching and eviction policies, fine-grained kernel preemption strategies, dynamic work-stealing schedulers, and programmable GPU kernel observability tools. \sys-based policies improve throughput by up to $4.8\times$ and reduce tail latency by up to $2\times$ compared to static heuristics, while instrumentation-only deployments incur low overhead. This demonstrates that cross-layer eBPF-based programmability spanning host and device is an effective abstraction for GPU-centric systems. In summary, our contributions are:

\begin{itemize}[leftmargin=*,itemsep=0pt]
  \item We design an extensible OS policy interface for GPU memory, scheduling, and observability.

  \item We implement \sys, an eBPF runtime with SIMT-aware verification spanning host driver and GPU device.

  \item We show \sys policies improve throughput and reduce tail latency across inference, training, and search workloads.
\end{itemize}

\section{Background and Motivation}

\subsection{The Diversity of GPU-Centric Workloads}
% \xiangyu{Workloads in CPU clusters}

\begin{figure*}[t]
    \centering
    \includegraphics[width=\textwidth]{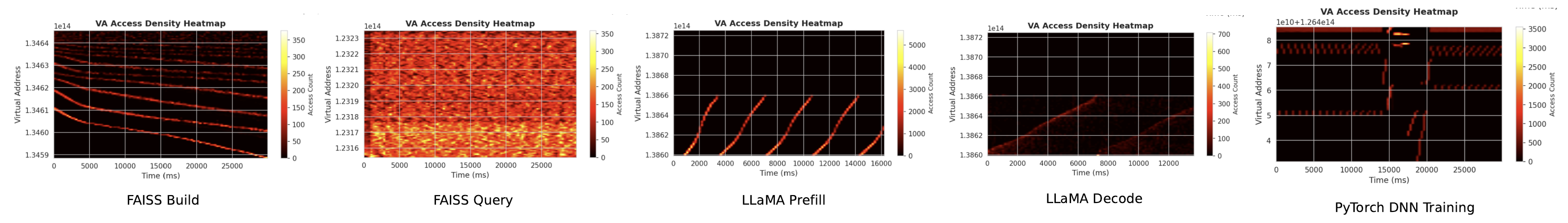}
    \caption{Memory access (page fault) patterns vary across GPU workloads: faiss Build exhibits sequential scans; faiss Query random accesses; llama.cpp MoE Prefill shows periodic sequential patterns while Decode exhibits sparse random accesses; PyTorch DNN shows periodic block accesses.}
    \label{fig:access_patterns}
\end{figure*}

\begin{figure}[t]
    \centering
    \includegraphics[width=\columnwidth]{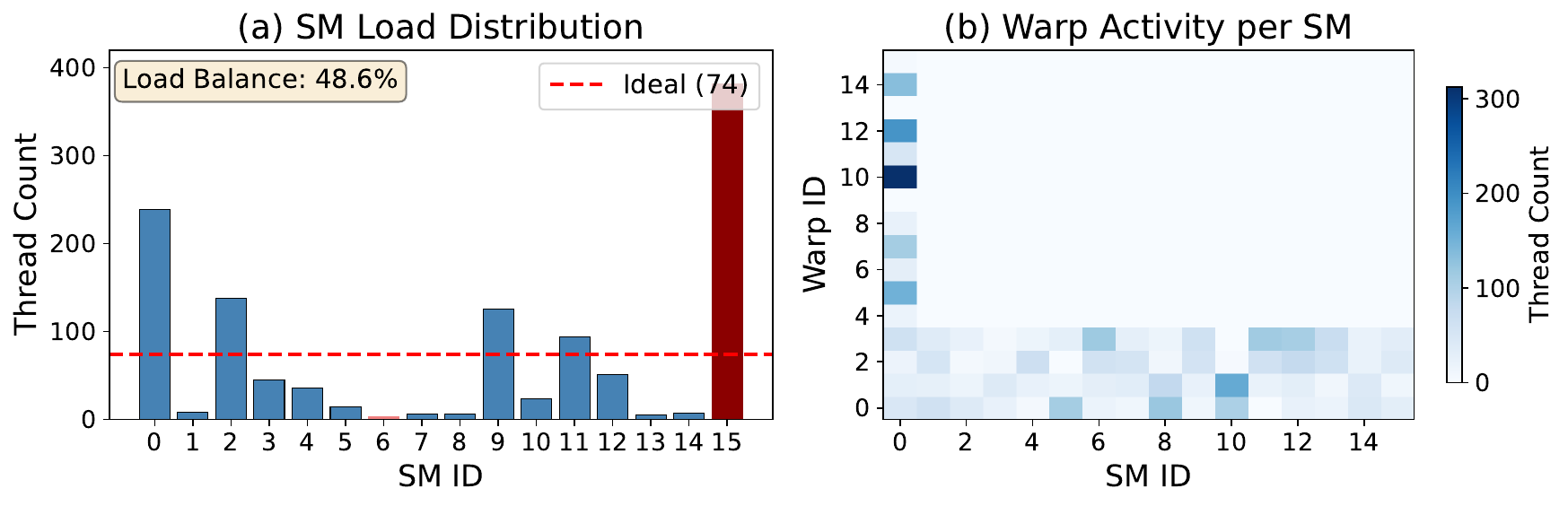}
    \caption{GPU thread scheduling imbalance observed via eBPF tracing. (a) SM load distribution shows 127$\times$ imbalance (SM~15: 382 threads vs SM~6: 3 threads). (b) Warp activity heatmap reveals SM~0 concentrates work in high-numbered warps while other SMs underutilize warp slots.}
    \label{fig:thread_scheduling}
\end{figure}

Modern GPU systems run diverse workloads with conflicting resource demands. Generative AI inference exhibits compute-intensive prefill, bandwidth-bound decoding~\cite{zhong2024distserve,patel2024splitwise}, irregular KV-cache offloading~\cite{sheng2023flexgen,kwon2023efficient}, and sparse expert activations~\cite{gale2023megablocks,he2021fastmoe}. Dense neural network training demands predictable synchronization, while GNN and embedding workloads exhibit irregular, pointer-chasing accesses resistant to prefetching. Vector search adds variability with random-access query and sequential indexing. These distinct memory behaviors (\Cref{fig:access_patterns}) and imbalanced thread scheduling across SMs (\Cref{fig:thread_scheduling}) are poorly served by static policies. In multi-tenant environments, latency-critical inference conflicts with throughput-maximizing training, creating contention that fixed partitioning cannot resolve.

% \xiangyu{The following might need a little bit of rewriting. I do not know the relationship with the previous sentence. Does it mean the previous one needs schedulers not to treat per-GPU as black box? or treat different GPUs differently?}
% \yusheng{I think we can remove them.}
% \xiangyu{It is OK to roughly mention these scenarios (because people might have already known). However, instead of purely saying what they are doing, we should give one central sentence: they all generate heterogeneous workload.}

\subsection{GPU Systems Architecture}

Modern GPU software stacks consist of three major layers:

% \begin{figure}[t]
%     \centering
%     \includegraphics[width=\columnwidth]{img/motivation_silos.pdf}
%     \caption{In modern GPU stacks, user-space applications have distinct requirements, but the OS driver is \emph{unprogrammable} and performs one-size-fits-all management. GPU device state remains \emph{invisible} to the host, preventing workload-aware policies.}
%     \label{fig:motivation_silos}
% \end{figure}

% need a better figure

% \xiangyu{High-level}
\paragraph{User-Space Logic (Applications \& Runtimes)}
% userspace include 2 parts: runtime and application. discuss them separately.
High-level frameworks (e.g., PyTorch) and runtimes (e.g., CUDA libraries) operate in user space, managing operation graphs, tensor layouts, and command submissions via memory-mapped interfaces or ioctl calls to drivers. These components encode rich semantic information, such as neural network structures and latency constraints. Runtime systems like vLLM~\cite{kwon2023efficient}, fine-grained sharing frameworks such as Salus~\cite{yu2020fine}, and UVM-focused runtimes like PILOT~\cite{ravi2021pilot} implement resource management policies at the application level.
%Therefore, we believe we should leverage existing software and utilize eBPF to relatively easily add some flexibility, thus achieving the goal of programmability.
% 

\paragraph{The One-size-fits-all Kernel Driver}
The GPU driver acts as the OS component for GPU systems, managing privileged low-level hardware mechanisms including the Memory Management Unit (MMU), hardware interrupts, and kernel scheduling. Typical drivers are monolithic, implementing application-agnostic resource-management algorithms such as Least Recently Used (LRU)-based page eviction, round-robin scheduling for context switching, and tree-based prefetching in Unified Virtual Memory scenarios~\cite{gu2020uvmbench,amd-oversubscription}. However, these application-agnostic policies provide no mechanism to adapt eviction, prefetching, or scheduling decisions based on workload characteristics or runtime conditions~\cite{gu2020uvmbench,ravi2021pilot}. They also lack multi-tenant coordination primitives—such as priority-based resource allocation or performance isolation between co-located workloads—that modern datacenter deployments require~\cite{wang2024gcaps,fan2025gpreempt}.
% \weichen{Complementing GPreempt, GCAPS~\cite{wang2024gcaps} demonstrates recent driver-level refactoring by implementing priority-based preemption for real-time guarantees. However, similar to prior approaches, GCAPS embeds a fixed scheduling policy within the driver, lacking a programmable interface for defining arbitrary resource management policies at runtime.}
% Updating drivers from scratch is extremely cumbersome, involving massive code updates, and could potentially introduce new bugs. Furthermore, much of the performance of driver software is application-scenario dependent, making it impossible to rewrite a driver for every new application scenario. 

\paragraph{Device-Side Execution}

The GPU device executes computations via user-defined device kernels and proprietary vendor firmware, which are compiled into static binaries specifying precise thread-level SIMT behaviors. GPU hardware schedulers manage thread execution at warp granularity. Device kernels and firmware lack runtime adaptability, making critical internal states (such as warp divergence, cache thrashing, and thread-level execution bottlenecks) opaque to host-side drivers. This architectural opacity constrains runtime workload-adaptive optimizations.

In summary, modern GPU software stacks include statically-defined policy logic embedded within monolithic kernel drivers, closed-source firmware, and opaque device kernels. This architectural coupling limits runtime policy flexibility, programmability, and adaptability in GPU-centric systems.

\subsection{Limitations of Existing Extensibility Mechanisms}

% \xiangyu{I think we can simplify the description of various related works by only referencing them but not mentioning their details.
% Probably, it is better to highlight limitations.}

Existing approaches face limitations:

\paragraph{Driver-Level Kernel-Space Policies.}

Recent research has introduced driver-level resource control by moving policy decisions into kernel drivers or intercepting driver-level events. Systems like TimeGraph~\cite{kato2011timegraph}, Gdev~\cite{kato2012gdev}, GPREEMPT~\cite{fan2025gpreempt}, LithOS~\cite{coppock2025lithos}, GCAPS~\cite{wang2024gcaps}, and kernel interception methods~\cite{kernelintercept} embed customized scheduling and memory-management logic into kernel-space or user-space driver modules. Despite improved performance, these approaches embed static policies requiring kernel modifications, restricting dynamic adaptation and safe deployment of new policies.

\paragraph{Host User-Space Runtimes and Libraries}
Recent frameworks such as Paella~\cite{ng2023paella}, Pie~\cite{gim2025pie}, XSched~\cite{shen2025xsched}, and KTransformers~\cite{chen2025ktransformers} expose programmability at the user-space layer. Paella provides software-defined GPU scheduling for model serving; Pie enables developers to customize inference loops and KV-cache management via ``inferlets''; XSched supports flexible, user-defined scheduling through its XQueue abstraction; and KTransformers exposes programmable hybrid CPU–GPU scheduling and expert-placement strategies. LithOS~\cite{coppock2025lithos} also re-implements runtime stacks with OS-like abstractions in user-space.

Despite this programmability, user-space systems share three limitations. 
First, programmability remains application-bound: developers must write policies for a specific framework or runtime, often requiring code changes or porting applications onto specialized stacks, preventing reuse across ecosystems.
Second, user-space runtimes lack global visibility and coordination across multi-tenant, multi-framework environments. Confined to isolated process silos, they cannot safely share memory or bandwidth, nor mitigate interference from co-located tenants. As a result, applications conservatively pre-allocate resources (e.g., GPU memory) at startup to avoid contention~\cite{xiao2018gandiva,yu2020fine}. Likewise, hardware partitioning such as NVIDIA's Multi-Instance GPU (MIG) imposes fixed resource boundaries that user-space runtimes cannot dynamically rebalance.
Third, user-space frameworks lack privileged access to low-level driver mechanisms, leaving policies constrained to coarse-grained GPU kernel launch boundaries. This precludes precise preemption for latency-sensitive tasks and prevents fine-grained compute–transfer overlap. Moreover, they cannot execute policy logic on latency-critical paths (e.g., page-fault handlers), nor exploit GPU-side warp-level or barrier-level access patterns.

\paragraph{Binary Instrumentation and Profiling Tools on Device}
\label{sec:background}
Tools such as NVBit~\cite{villa2019nvbit} and Neutrino~\cite{neutrino} inject logic directly into GPU binaries. NVBit supports dynamic SASS-level instrumentation, while Neutrino provides instruction-level profiling via programmable assembly-layer probes with an eBPF-like interface. Despite high kernel visibility, these tools face three limitations. First, they lack safety guarantees, as injected logic can cause crashes or deadlocks. Second, they target offline profiling rather than online policy enforcement, operating in read-only modes that cannot control memory placement or thread scheduling. Third, lacking host-side integration, they cannot coordinate policies across processes. Additionally, assembly-level instrumentation restricts expressiveness and portability~\cite{neutrino}. Our prior workshop paper eGPU~\cite{egpu} extends eBPF to GPU instrumentation but remains limited to observability with prohibitive overhead (see \S\ref{sec:eval}).

\paragraph{Host-Based Kernel Programmability (CPU eBPF)}
CPU-side eBPF has been used to customize OS scheduling and resource management (e.g., \texttt{sched\_ext}, \texttt{cache\_ext}). However, extending this model to GPUs faces two barriers. First, host-based eBPF treats the GPU as a black box: it can trace driver commands but cannot observe internal execution state (e.g., warp divergence), preventing reaction to on-chip bottlenecks. Second, unlike the Linux CPU scheduler refactored to support \texttt{sched\_ext}, monolithic GPU drivers expose no safe programmable extension points. Their resource management mechanisms are tightly coupled to low-level hardware interactions (e.g., MMU invalidation), and naïvely exposing such controls risks resource leaks, hardware hangs, or kernel panics.
\section{Design Principles}

\sys follows three design principles:

\paragraph{Principle 1: OS-level, Transparent and Dynamically Programmable Interface.}
GPU resource management policies should reside at an OS-level interface, rather than within vendor drivers or application-specific runtimes. Policies must be dynamically separable from mechanisms, runtime-updateable without application restarts, and enforceable across unmodified GPU workloads.

\paragraph{Principle 2: Cross-layer host–device policy abstraction.}
Host and GPU devices should not be managed as separate execution domains. Instead, the system should provide a unified programming abstraction and shared control plane across CPU and GPU, allowing consistent policy decisions (e.g., memory placement, scheduling).

\paragraph{Principle 3: SIMT-aware static safety and efficiency.}
Device-side policies must align with GPU SIMT execution semantics. The system should perform static, load-time verification and optimization, ensuring warp-uniform control flow, bounded resource usage, memory safety, and bounded overhead.

\section{\sys Design}

\subsection{Challenges}
\label{sec:challenges}

We elaborate on the three challenges introduced in \S1.

\subsubsection{C1: Lack of a Safe and Expressive OS-Level GPU Policy Interface}
GPU resource management needs to run at the driver layer to be effective, but GPU drivers were not designed to expose a programmable interface. Exposing low-level mechanisms (page tables, command buffers, interrupt handlers) to eBPF would give policies expressiveness but compromise driver stability and tenant isolation. Conversely, constraining programmability to high-level abstractions ensures safety but limits the expressiveness needed for complex memory placement and scheduling decisions. Thus, the challenge lies in carving out a narrow, stable, and verifiable GPU policy interface: expressive enough to implement policies, yet constrained enough to prevent buggy policies from corrupting device state or crashing the kernel.

\subsubsection{C2: Mismatch Between CPU Extension Semantics and the GPU SIMT Execution Model}
Applying existing CPU-oriented eBPF mechanisms to GPU device is challenging due to semantic mismatches with the GPU SIMT execution model. CPU-side eBPF assumes single-threaded execution, whereas GPU kernels execute in warps of 32 threads that must follow warp-uniform control paths. Running scalar eBPF logic per thread introduces warp divergence, serialization overhead, and deadlock risks from thousands of concurrent memory operations unchecked by the CPU-side verifier. Furthermore, GPUs lack isolation and recovery points, meaning an unbounded loop or invalid access in policy logic can halt the entire device. This semantic mismatch makes reuse of CPU-oriented eBPF inefficient and unsafe.

\subsubsection{C3: Absence of Efficient Mechanisms for Host–Device Shared State}
GPU policies require coordinated host–device resource management, which today lacks efficient and consistent abstractions for shared policy state. GPU-accessible host memory exhibits orders-of-magnitude higher latency than GPU global memory, while GPU memory is capacity-limited and optimized for throughput. Meanwhile, CPU and GPU components operate asynchronously at different timescales, causing policy decisions to rely on stale or inconsistent state unless synchronized.

\subsection{High-Level Architecture}
\label{sec:arch}

\begin{figure}
\centering
    \includegraphics[width=\columnwidth]{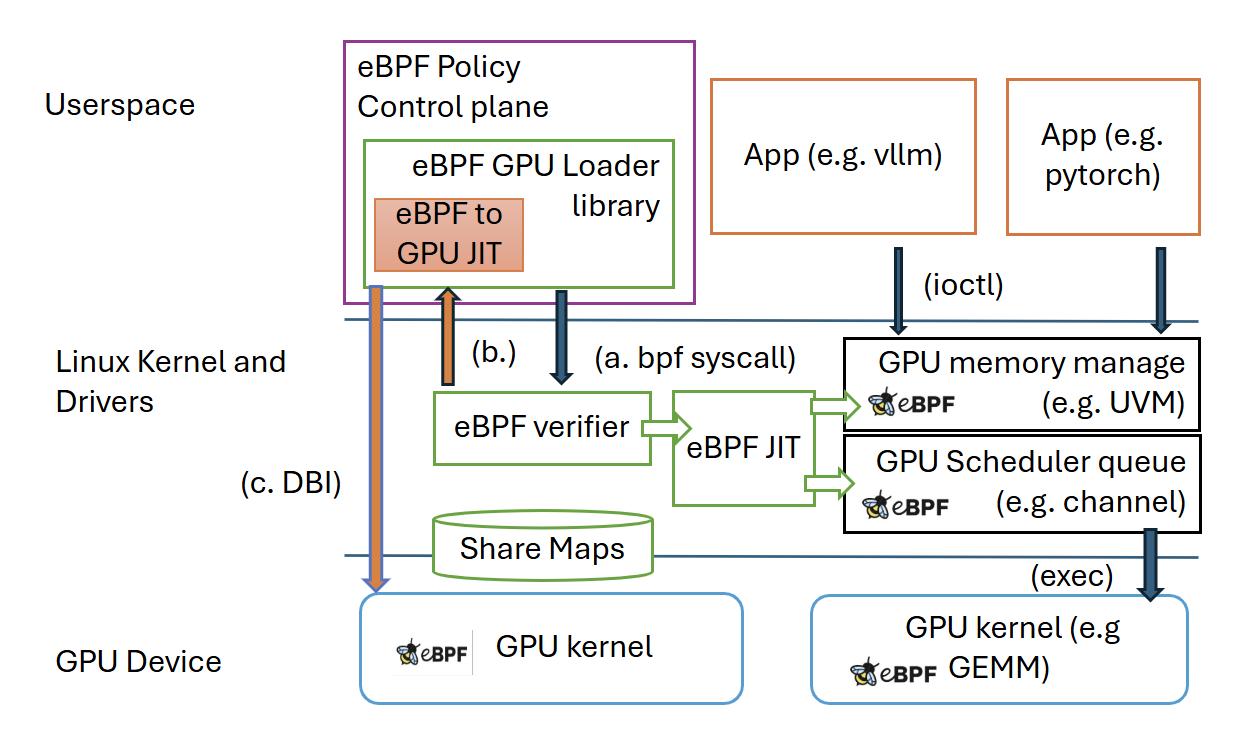}
\vspace{-.5ex}
\caption{\sys architecture: cross-layer eBPF runtime spanning kernel driver and GPU device. The control plane deploys policies via (a) bpf syscall to verifier and driver hooks (b) GPU JIT after verification; (c) DBI injects trampolines into GPU kernels. Shared maps enable state coordination across layers.}
\label{fig:system-design}
\end{figure}

Figure~\ref{fig:system-design} shows the architecture of \sys.

% \sys exposes \emph{hierarchical cross-layer maps}: logical eBPF map abstractions whose backing storage is automatically partitioned and replicated across host DRAM, GPU global memory, and optionally per-SM scratch space, with consistency and placement managed transparently by the runtime. Policy developers access maps via standard eBPF helpers regardless of execution context, without manual DMA transfers, memory placement, or address translation.

\paragraph{User-space Control Plane.}
Policy authors use standard eBPF tooling (e.g., \texttt{clang/libbpf}, \texttt{bpftool}) to build policy bytecode. The control-plane links against a loader library that installs policies and configures eBPF maps. It enables runtime policy redeployment and reconfiguration (e.g., eviction thresholds, sampling rates, priorities) without application or kernel restarts, and exports runtime metrics (e.g., region hotness, tenant latencies).

\paragraph{Host Driver and Device Runtime.}
\sys provides a unified runtime across host driver and GPU device contexts, managing verification, compilation, and map placement. It employs a SIMT-aware verifier (\S\ref{sec:verification}) that extends standard eBPF checks to verify warp-uniformity, SIMT control flow, and GPU synchronization constraints. Verified programs are JIT-compiled into native host code or GPU-compatible instructions (e.g., PTX). The runtime manages hierarchical logical eBPF map abstractions whose backing storage is automatically partitioned and replicated across host and device with consistency.
Policy execution occurs at structured hook points defined in GPU drivers and device. On the host, \sysdriver exposes \texttt{struct\_ops}-style hooks (e.g., region activate/eviction, GPU kernel scheduling). Host policies return concise policy decisions enforced via trusted kfuncs. On the GPU device, \sys injects trampolines at hook points and invokes device-side handlers at warp granularity. Device handlers update cross-layer maps with performance signals and make control decisions (e.g., prefetch, block-level scheduling) via trusted helpers that the runtime enforces.

\subsection{\sys{} Policy Interfaces}
\label{sec:design-interfaces}

\sys introduces three eBPF program types: two in the driver‑level ( \texttt{BPF\_PROG\_TYPE\_GPU\_MEM} for memory‑placement policy and \texttt{BPF\_PROG\_TYPE\_GPU\_SCHED} for scheduling decisions), and one in the device‑level ( \texttt{BPF\_PROG\_TYPE\_GPU\_DEV} for GPU kernel instrumentation).
These program types coexist with existing kernel eBPF program types and have different SIMT‑specific verification rules.

\subsubsection{Memory Interface}
\label{sec:design-mem}

\sys treats memory placement as a programmable cache. It exposes four events (\emph{activate}, \emph{access}, \emph{evict\_prepare}, and \emph{prefetch}) and allows host and device policies to cooperate. \sys regions correspond to GPU hardware-managed units, such as 2MB physical UVM chunks or VA blocks, and finer-grained 4KB pages. This granularity balances policy precision against metadata overhead. Using the regions abstraction, \sys hides vendor-specific and hardware complexities. The host driver exports these events through a \texttt{struct\_ops} table. Handlers encode decisions through an enum field in the context and may reorder but never remove regions in the kernel‑maintained eviction list. The kernel enforces safety and correctness, including fallback FIFO eviction under pressure. Policies maintain their own metadata in maps and express placement and prefetch decisions through these handlers.

Device policies can be triggered at given memory operations and receive a compact warp‑uniform context. Operations like prefetch can be performed on device and then trigger host-side prefetch handlers, enabling the host to combine driver information with device‑side predictive logic. The following interface illustrates the host and device memory policy hooks:

\begin{lstlisting}
// Host-side policy handlers for GPU memory management
struct gpu_mem_ops {
  // Region created or eligible for device placement
  int (*gpu_activate)(gdrv_mem_add_ctx_t *ctx);
  // Memory subsystem observes region use;
  // decide cache hit/eviction policies
  int (*gpu_access)(gdrv_mem_access_ctx_t *ctx);
  // Region about to leave device-resident working set
  int (*gpu_evict_prepare)(gdrv_mem_remove_ctx_t *ctx);
  // Safe points (e.g. faults) for prefetch
  int (*gpu_prefetch)(gdrv_mem_prefetch_ctx_t *ctx);
};

// Host-side kfuncs: reorder regions in eviction list
void bpf_gpu_move_head(gdrv_mem_list_t *list,
                               gmem_region_t *r);
void bpf_gpu_move_tail(gdrv_mem_list_t *list,
                               gmem_region_t *r);

// Device-side memory handler interface
struct gdev_mem_ops {
    // Warp observed memory access;
    void (*access)(gdev_mem_access_ctx_t *ctx);
    // GPU kernel fence point
    void (*fence)(gdev_mem_fence_ctx_t *ctx);
};

// Device-side helpers
// Request prefetch, triggers handler in host driver
 int gdev_mem_prefetch(gmem_region_t* r);
\end{lstlisting}

\subsubsection{Scheduling Interface}
\label{sec:design-sched}

\sys exposes scheduling policy control at two levels: hardware‑queue lifecycle (host‑side) and thread‑block scheduling (device‑side). The host participates in GPU queue admission and priority control through policies. Policies set queue priority, timeslice, and may reject or defer queue creation. Properties are written directly into firmware‑visible structures, ensuring enforcement beyond user‑space hints. Policies may also invoke the \texttt{gdrv\_sched\_preempt} kfunc to trigger cooperative preemption through the driver's native context‑switch mechanism.

Within the GPU, \sys provides a work‑stealing thread‑block scheduler. GPU kernels expose their logical work units; persistent worker blocks select and process units while device‑side eBPF handlers steer scheduling decisions via \texttt{gdev\_block\_ctx}. The verifier ensures that scheduling decisions respect GPU kernel invariants (e.g., selecting only local work units). \sys can inject this scheduler via binary rewriting or allow developers to integrate it.

\begin{lstlisting}
// Host-side policy handlers for GPU hardware
// queue scheduling
struct gpu_sched_ops {
  // Hardware queue creation; set prio/timeslice,
  // return 0 to accept or <0 to reject/defer
  int  (*task_init)(gsched_queue_ctx_t *ctx);
  // Hardware queue destruction; cleanup policy
  void (*task_destroy)(gsched_queue_ctx_t *ctx);
};

// Host-side kfuncs
void bpf_gpu_set_attr(gsched_queue_ctx_t *ctx, u64 us);
void bpf_gpu_reject_bind(gsched_queue_ctx_t *ctx);

// Device-side block scheduling handlers
struct gdev_sched_ops {
  // Worker block starts/exits new work unit
  void (*enter)(gdev_block_ctx_t *ctx);
  void (*exit)(gdev_block_ctx_t *ctx);

  // Device function starts/ends
  void (*probe)(gdev_block_ctx_t *ctx);
  void (*retprobe)(gdev_block_ctx_t *ctx);

  // Return whether to steal work (TB scheduler)
  bool (*should_try_steal)(gdev_block_ctx_t *ctx);
};
\end{lstlisting}

\subsection{Runtime Verification and Optimizations}
\label{sec:verification}

\sys policies execute directly within privileged GPU driver and kernel contexts, making safety guarantees critical. We adopt a standard eBPF threat model: we trust system administrators loading policies, but consider policy code potentially buggy or misconfigured. The trusted computing base (TCB) includes the kernel and \sys kernel module, GPU compiler backend, and GPU driver/firmware. Policy authors only interact with eBPF and do not write native device code. For host-side verification, we reuse the existing Linux kernel eBPF verifier unchanged, relying on standard type-checking, bounded-loop, and memory-safety constraints. \sys-specific kfuncs and hooks are declared via BPF Type Format (BTF) metadata.

\subsubsection{SIMT-aware Verification}

To solve the SIMT mismatch described in C2, \sys introduces a SIMT-aware static verification model enforcing warp-level uniformity constraints to ensure efficient and safe GPU-side policy execution.
The key insight is that we distinguish \emph{warp-uniform} values (identical across warp threads, e.g., region identifiers) from \emph{lane-varying} values (thread-local GPU state). Policies may compute freely using lane-varying values, but these must not influence control-flow decisions or directly produce side effects visible outside a warp. To guarantee warp-level coherence, \sys mandates that branch conditions, loop bounds, and map-update keys be warp-uniform or derived via warp-level aggregation. Additionally, \sys prohibits GPU-wide barriers, global synchronization primitives, and non-uniform atomic operations, and imposes resource budgets per policy hook to bound memory and thread resource usage. These constraints ensure safe policy integration that respects GPU hardware execution semantics.

\subsubsection{Warp-level Execution Optimizations}

Direct scalar execution of eBPF after SIMT verification on GPU threads still causes resource overhead (e.g. duplicate memory update to maps) due to the SIMT. To resolve this, \sys introduces a warp-level aggregated execution model, executing policy logic once per warp using a designated "warp leader" (typically lane 0). Each hook invocation conceptually follows a two-phase approach: each lane independently computes lane-local contributions (e.g., effective addresses, counters) without influencing control flow or directly updating shared state. These contributions are then aggregated at warp granularity. Subsequently, the warp leader executes the policy handler once using aggregated inputs and warp-uniform context. The resulting decision is broadcast back to all lanes. This design minimizes GPU overhead while preserving eBPF's scalar semantics. To further reduce runtime overhead, \sys employs compiler-level optimizations (specialization and inlining) tailored for warp-level execution.

\subsubsection{Hierarchical Cross-layer Maps for State Coordination}

GPU policy decisions require coherent sharing of policy state across asynchronous CPU and GPU contexts. To transparently bridge CPU-GPU memory hierarchies, \sys introduces cross-layer maps. Logically, each map is a single unified key-value store accessible from host-side driver hooks, GPU-side device policies, and user-space control planes, and verifiable through eBPF's existing model. Physically, maps are realized as hierarchical shards across host DRAM, GPU global memory, and GPU-local storage (e.g., shared memory, SM-local). The runtime dynamically determines shard placement based on access frequency, latency, and capacity constraints. Following eBPF's soft-state philosophy, \sys maps provide relaxed, eventual consistency. Instead of strong global ordering, our runtime periodically merges GPU-local map shards into canonical snapshots at synchronization points, such as GPU kernel completion boundaries. This snapshot-based model is sufficient for policy decisions relying on approximate statistics. Occasional staleness affects decision optimality but cannot violate correctness invariants such as memory integrity, which remain enforced by the GPU driver and hardware MMU.

\section{Implementation}
\label{sec:implementation}

\subsection{Components Overview}

We implemented \sys on Linux 6.15, extending the kernel with a standalone file ($\sim$1 KLOC) that handles hook registration and verifier extensions. We integrated lightweight instrumentation ($\sim$100 LOC) into NVIDIA's open GPU kernel modules to expose memory management and scheduling decision points (\S\ref{sec:impl-host}). Our userspace loader and JIT infrastructure ($\sim$10 KLOC) builds upon bpftime's framework, while an LLVM-based backend ($\sim$1 KLOC) translates the restricted \sys eBPF instruction subset into GPU device code (PTX) (\S\ref{sec:impl-gpu}).

\subsection{Host-side Integration}
\label{sec:impl-host}

\sys extends the NVIDIA Open GPU Kernel Modules~\cite{nvidia-open-gpu}, instrumenting the UVM module at page fault handler and prefetch logic. At each hook, the driver constructs a context containing fields (e.g., region range aligned to 2MB GPU physical chunks, VA blocks or 4KB pages, fault address, block size) and invokes an attached eBPF handler. Currently, eviction hooks operate at 2MB block granularity, while prefetch hooks operate at page granularity. Each handler returns a decision to execute or bypass default kernel logic. The kernel maintains a doubly-linked eviction list; handlers may reorder regions via provided helpers, but the kernel retains eviction authority under memory pressure.

We instrument Time-Slice Group (TSG) lifecycle events, including initialization (\texttt{task\_init}) and teardown (\texttt{task\_destroy}). The eBPF policy can configure scheduling parameters such as TSG priority, timeslice duration, and runlist interleave frequency via the \texttt{bpf\_gpu\_set\_attr} kfunc, and can reject task binding via the \texttt{bpf\_gpu\_reject\_bind} kfunc. Our kernel module extends the Linux eBPF verifier by registering GPU-specific struct\_ops and kfunc interfaces, leveraging BTF annotations to enforce type and memory safety.

\subsection{GPU Runtime}
\label{sec:impl-gpu}

After the Linux kernel verifies a \sys device eBPF program, our loader forwards bytecode and BTF metadata to a userspace backend, translating the restricted eBPF subset into GPU PTX code. We reuse Linux's eBPF verifier to enforce standard memory safety, bounded loops, and type correctness. An additional verifier pass leverages BTF metadata annotations that mark warp-uniform fields: at load time, our verifier performs dataflow analysis on eBPF bytecode, propagating warp-uniformity through arithmetic, logical, and memory operations. Control-flow constraints are enforced by traversing the policy's CFG to reject lane-varying dependencies in branch predicates or loop bounds. Map-update instructions must use warp-uniform keys derived from uniform registers or warp-level reductions. GPU-wide barriers, global synchronization primitives, and non-uniform atomics are disallowed by pattern-matching. Each hook type carries resource budgets limiting instructions, helper invocations, and memory operations. 

Before each GPU kernel runs, \sys dynamically intercepts CUDA runtime APIs to extract GPU kernel PTX, rewrite it with binary trampolines, and load instrumented GPU kernels back without recompilation or restarting the application. These trampolines, placed at GPU kernel entry, selected memory instructions (such as global loads and atomic operations), and execution-phase boundaries, insert prologue/epilogue logic for SIMT register preservation. In each trampoline, per-lane inputs are aggregated using \texttt{\_\_ballot\_sync} and \texttt{\_\_shfl\_sync}, with a warp leader selected via \texttt{\_\_ffs(\_\_activemask())} to execute the scalar eBPF handler once per warp, broadcasting results via shuffle instructions; verified eBPF code executes after kernel launch, supported by pre-compiled GPU helper functions. During JIT compilation, we inline helper functions and map accesses to reduce call overhead. For scheduling, \sys implements policies using persistent GPU workers as long-lived kernels that poll and claim logical work units; on Blackwell GPUs, we use cluster launch control APIs to schedule thread blocks. For memory management instrumentation, \sys attaches eBPF policy at memory instructions, GPU kernel function boundaries, and tracepoints to issue policy hints to the host driver. Unlike our prior work eGPU~\cite{egpu} which relies on atomic synchronization, \sys{} avoids cross-SM synchronization primitives to prevent hardware stalls.

\paragraph{Hierarchical Maps and Cross-Domain State}

We use eBPF maps in GPU-accessible memory as a shared state abstraction between host-driver and GPU-device contexts. Map shards are placed according to access patterns: cold global state resides in host DRAM and is periodically refreshed from GPU-side snapshots; frequently accessed state resides in GPU global memory and is asynchronously flushed to host; hot per-warp or per-SM data is cached in GPU shared memory for low latency. Consistency across shards is maintained via snapshot-based aggregation at GPU kernel completion boundaries. A runtime daemon asynchronously flushes GPU-local shards to host-visible canonical map instances, providing coherent snapshots to host-side policies without synchronization overhead. Shard updates occur strictly at warp granularity through warp-leader threads without GPU atomic operations. For example, GPU-side handlers accumulate per-region access counters in warp-local registers, aggregate them per warp, and store results into GPU-local shards. The periodic snapshot flush transfers these shards to the host for policy decisions.
\section{Evaluation}
\label{sec:eval}

We evaluate \sys using three research questions:

\begin{itemize}[leftmargin=*]
  \item \textbf{RQ1 (Single-Tenant Management):} How much performance gain can \sys's programmable memory and scheduling policies provide on oversubscribed single-tenant workloads?
  \item \textbf{RQ2 (Multi-Tenant Management):} Does \sys's cross-layer policy runtime improve tail latency, throughput, and resource fairness compared to user-space and global policies in multi-tenant settings?
  \item \textbf{RQ3 (Programmability \& Mechanism Overhead):} Is \sys sufficiently programmable to support diverse policies, and what is the overhead of its core mechanisms and observability capabilities?
\end{itemize}

These questions cover single-tenant optimization (RQ1), multi-tenant coordination (RQ2), and system costs (RQ3).

\subsection{Methodology and Setup}

We evaluate \sys on two machines: Server~A with Intel Core Ultra~9 285K (24 cores), 128~GB DDR5 RAM, and NVIDIA RTX~5090 GPU; Server~B with dual Intel Gold 6138 (80 cores), 256~GB RAM, and NVIDIA P40 GPU. Unless otherwise specified, experiments run on Server~A, averaging 10 trials using geometric mean. We compare against default UVM, UVM with user-space hints, native GPU scheduler, and framework-managed offloading where applicable. We use PyTorch (version 2.9.0) and vLLM (version 0.11.0) with custom allocator, llama.cpp (version 7101), and Faiss (version 1.13.0) with UVM build config for our evaluation. All tests with \sys policies require no application modifications.

\subsection{RQ1: Single-Tenant Memory and Scheduling Policies}

We evaluate \sys's memory policy interface on oversubscribed single-tenant setups where memory placement and prefetching dominate performance.

\subsubsection{Microbenchmarks}

\paragraph{Memory Policy Microbenchmarks.}
We evaluate host-device prefetch coordination using a modified vector-add GPU kernel~\cite{nvidia_cuda_samples} with stride access pattern and 40GB working set (1.25$\times$ oversubscription). Device-side L2 prefetch instructions (\texttt{prefetch.global.L2}) trigger non-blocking page faults, while host-side callbacks extend prefetching to additional pages. With device-only prefetch and default host prefetch at thread entry, we achieve 1.34$\times$ speedup; combined host-device stride-based eBPF prefetch achieves 1.77$\times$. However, sequential prefetch degrades performance by 8\% due to pattern mismatch, demonstrating that \sys's cross-layer design enables workload-aware prefetching where policy selection matters.

\begin{figure}
\centering
\includegraphics[width=0.9\columnwidth]{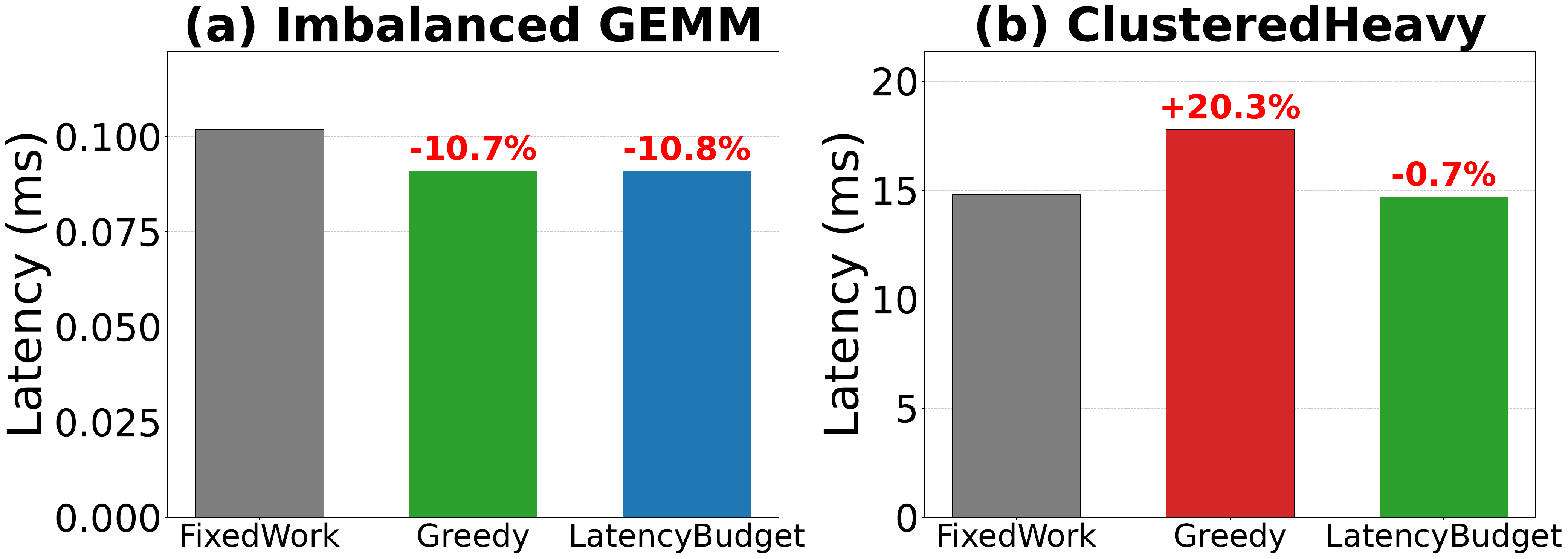}
\caption{\sys block-scheduling policies across workload regimes. No single policy dominates: always-steal works well under moderate imbalance (a) but becomes pathological under clustered heavy tails (b), where LatencyBudget matches baseline performance.}
\label{fig:clc-policies}
\vspace{-2ex}
\end{figure}

\paragraph{Single-Tenant Block Scheduler.}
We apply a \sys-based block scheduler driven by the GPU-side interface (Section~\ref{sec:design-sched}) to workloads with load imbalance. Using cluster-style launch where persistent worker blocks pull work units via \texttt{should\_try\_steal}, we evaluate three policies: FixedWork (static block assignment, no scheduler), Greedy (always-steal), and LatencyBudget (workload-aware with per-block time limits).
\Cref{fig:clc-policies} compares these policies across two GEMM workloads. Under moderate imbalance (\Cref{fig:clc-policies}a), both Greedy and LatencyBudget reduce latency by $\sim$11\%, as always-steal scheduling reclaims wasted tail time. Under heavy-tail workloads where 10\% of blocks perform 100--200$\times$ more work (\Cref{fig:clc-policies}b), Greedy increases latency by 20\% due to contention on heavy blocks. LatencyBudget, which caps per-block stealing time, matches fixed-work baseline. This highlights \sys's programmability: different workloads require different scheduling strategies.

\subsubsection{Case Studies}

\begin{figure}
\centering
\includegraphics[width=\columnwidth]{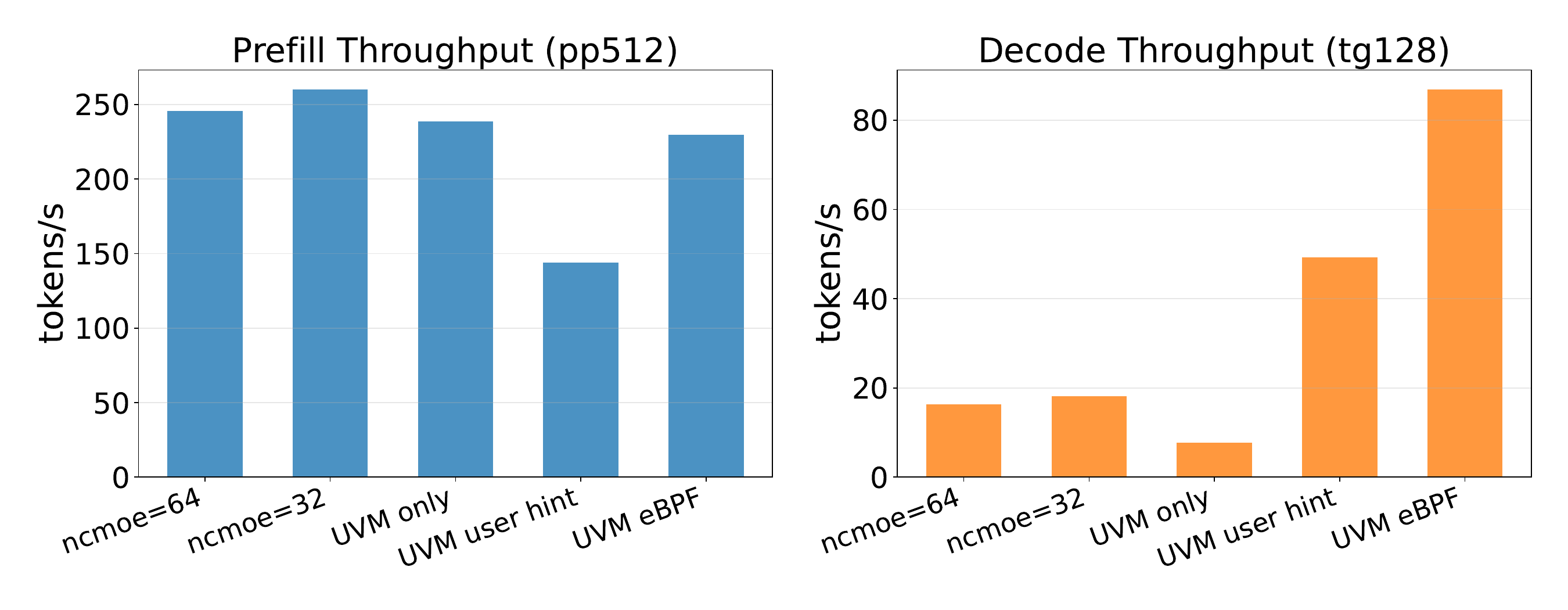}
\vspace{-1ex}
\caption{Prefill (pp512) and decode (tg128) throughput for GPT-OSS-120B MoE (59~GiB) on RTX 5090 (32GB). \sys eBPF prefetching achieves 4.8$\times$ speedup on decode (memory-bound) over framework offloading while maintaining competitive prefill performance.}
\label{fig:llama-expert-offload}
\vspace{-2ex}
\end{figure}

\paragraph{Expert Offloading (llama.cpp GPT-OSS-120B).}
We evaluate \sys on GPT-OSS-120B MXFP4 MoE (59~GiB, 116.83B parameters) in llama.cpp on RTX 5090 (32GB). The model requires 1.84$\times$ oversubscription, necessitating CPU-GPU memory tiering. We compare five configurations: framework-managed CPU offloading (\texttt{ncmoe=32/64}), default UVM, modify application to add user-space hints (\texttt{cudaMemAdvise}) with UVM, and \sys with eBPF-based expert prefetching.
\Cref{fig:llama-expert-offload} shows prefill and decode throughput. For decode (memory-bound), \sys achieves \textbf{4.8$\times$ speedup} over framework offloading. For prefill (compute-bound), framework offloading achieves 13\% higher throughput, but decode dominates end-to-end latency, so the 4.8$\times$ decode improvement outweighs the 13\% prefill overhead. Static hints can affect default LRU algorithms to pin most of the memory on GPU to improve performance, but still falls behind our policy, confirming that UVM requires new algorithms. The key insight is that MoE workloads exhibit predictable stride patterns during weight access and non-uniform page-level access frequency. \sys uses stride prefetch to overlap data transfer with computation, and LFU eviction to retain frequently-accessed pages. Unlike framework-level offloading that often migrates experts as atomic units, \sys operates at page granularity, enabling finer-grained compute-transfer overlap.

% \xiangyu{Are those hints easy to find in many scenarios? Maybe we should clarify it. Otherwise, it seems that \sys{} uses some tricks to win the competition.}

\paragraph{KV-cache Offloading (vLLM Qwen-30B MoE).}
We evaluate on the Qwen-30B FP8 MoE model ($\approx$30GB) on RTX 5090 (32GB). The model alone nearly fills GPU VRAM, so default no-offload configuration OOMs under load. We stress KV-cache growth by sending 100 concurrent requests from ShareGPT\cite{sharegpt} (single-round, no prefix caching), resulting in total memory footprint of 36--40GB (configured KV-cache at $\approx$60K tokens). Existing solutions typically offload either KV-cache or MoE experts, but not both; in contrast, UVM attempts on-demand migration of both, leading to thrashing.

We compare vLLM's CPU-offload mode (\texttt{--cpu-offload-gb 8}) against \sys with UVM and a sequential prefetch policy. Under memory pressure, the key challenge is managing two competing data types: KV-cache and model weights. KV-cache exhibits temporal locality (recent tokens accessed more frequently during attention) with per-request spatial locality, while expert weights show stride patterns during matrix operations. \sys applies prefetching with adaptive aggressiveness based on PCIe utilization and memory region (stride-based for weights, sequential for KV-cache), and LFU eviction to retain frequently-accessed pages from both types. This prevents mutual thrashing that occurs with UVM's default LRU when KV-cache growth displaces hot weight pages or vice versa.

\begin{figure}
\centering
\includegraphics[width=\columnwidth]{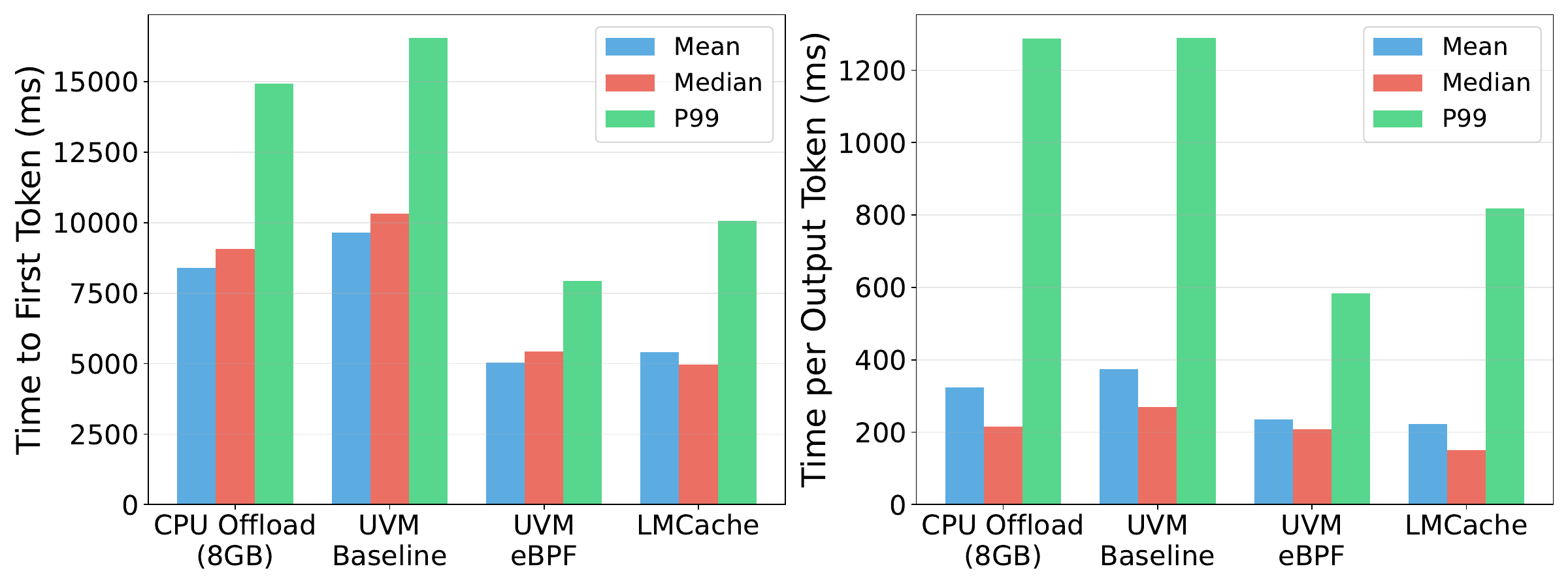}
\vspace{-1ex}
\caption{Time-to-first-token and decoding throughput on Qwen-30B FP8 MoE with 100 concurrent requests ($\approx$60K tokens KV-cache). \sys KV-aware sequential prefetch improves mean and p99 TTFT by 1.7--2$\times$ and decoding throughput by 1.3$\times$ over vLLM CPU-offload.}
\label{fig:vllm-kv-offload}
\vspace{-2ex}
\end{figure}

\sys improves mean and p99 time-to-first-token by 1.7--2$\times$ and decoding throughput by 1.3$\times$ over vLLM's default framework-managed offload (\Cref{fig:vllm-kv-offload}), matching the state-of-the-art KV cache offload framework LMCache~\cite{lmcache} with better tail latency. Notably, UVM without \sys policies performs worse than vLLM's offload, while \sys is 2--3$\times$ faster, showing driver-level policies can outperform framework offloading with better tail latency.

\begin{figure}
\centering
\includegraphics[width=\columnwidth]{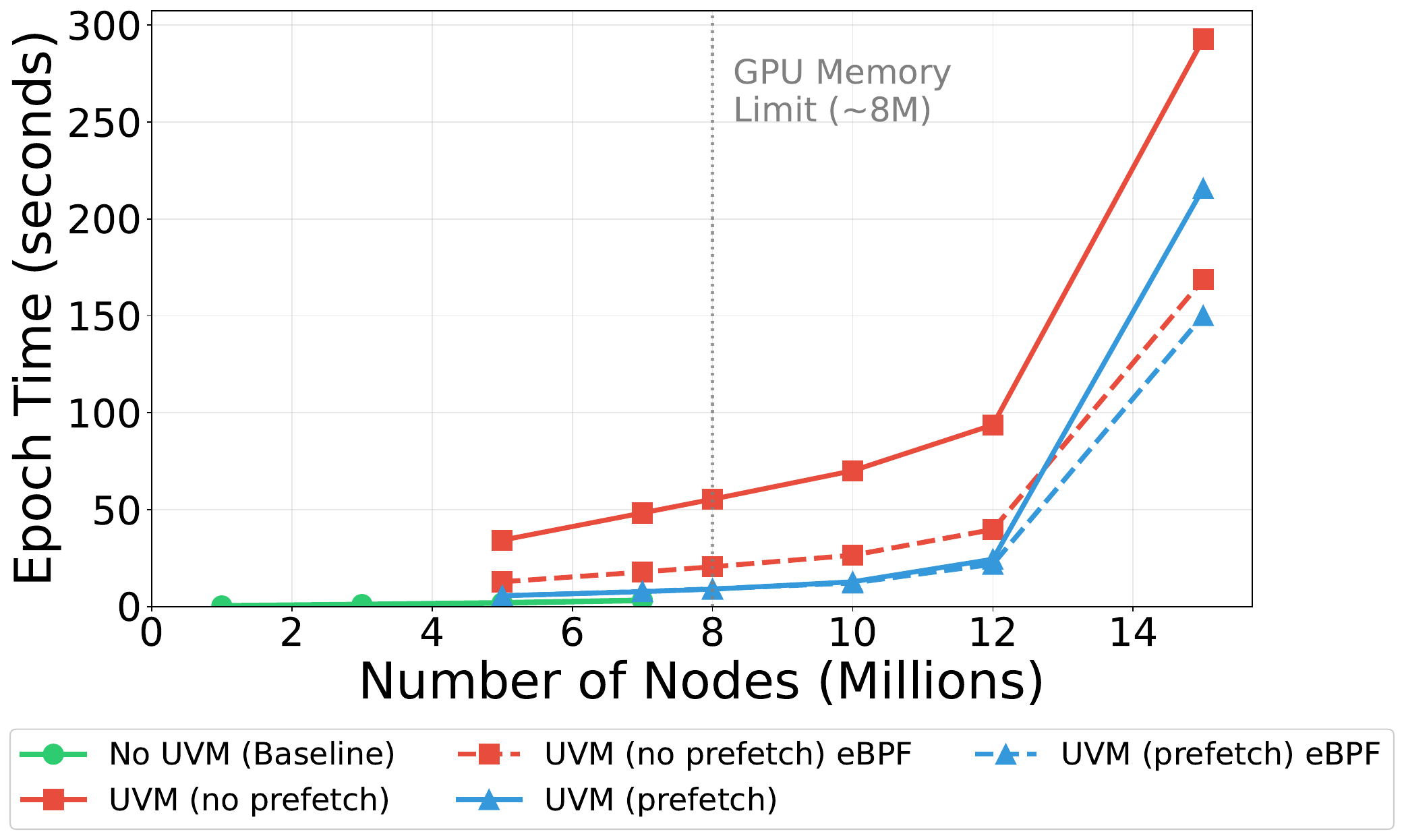}
\vspace{-1ex}
\caption{GCN training epoch time vs graph size. Native GPU allocation (green) OOMs beyond memory capacity. User-space prefetching (blue) eliminates page faults but requires application modification; \sys eBPF prefetching (red) achieves similar benefits transparently. Combined approaches (dashed blue) provide additional benefits under severe oversubscription.}
\label{fig:gnn-epoch}
\end{figure}

\paragraph{Graph Neural Networks (GNN Training).}
We evaluate \sys on PyTorch-based GCN training with random graphs of varying sizes (1M--15M nodes, 10 edges/node). \Cref{fig:gnn-epoch} shows epoch time across five configurations: native GPU allocation (no UVM), UVM with and without user-space prefetching (\texttt{cudaMemPrefetchAsync}), and each UVM configuration with \sys eBPF-based optimization. 

With UVM enabled via a custom allocator, each epoch allocates memory on CPU and migrates to GPU on demand, incurring overhead. \sys uses sequential prefetch with uprobe to trace PyTorch memory allocations and prefetch in advance, using history patterns to determine prefetch range.
Users can also express prefetching via \texttt{cudaMemPrefetchAsync} to eliminate page faults by migrating pages before GPU access, achieving 5.5$\times$ speedup at moderate oversubscription, but this requires application modification. When user-expressed prefetching is disabled, eBPF optimizes page fault handling by aggressive prefetching, achieving 2.65$\times$ speedup without modifying applications. Even with user-expressed prefetching, runtime page faults and eviction occur under memory pressure, and \sys's combined approach achieves 1.44$\times$ additional speedup by prefetching larger regions and reducing blocking faults.
Native GPU allocation (no UVM) achieves the highest performance when data fits in memory (1.89~s at 5M nodes) but fails with OOM beyond $\sim$8M nodes. UVM with \sys prefetching enables training at 15M nodes (2.17$\times$ oversubscription) with acceptable overhead. 

\paragraph{Faiss Vector Search.}
We evaluate \sys on vector similarity search using the SIFT dataset with an IVF4096,Flat index, scaling from 20M (9.5GB, fits in memory) to 50M (24GB, near boundary) and 100M (48GB, exceeds 32GB GPU memory). IVF indexes have a two-level structure: frequently-accessed centroids and larger posting lists. For oversubscribed datasets, \sys prefetches posting lists sequentially rather than UVM's default address-tree-based policy; device-side can also trigger prefetching of corresponding posting lists. For index construction, K-means iterations produce sequential scans that \sys detects and prefetches using stride prediction. \sys reduces build time by 21--29\%, with benefits scaling from 27\% to 40\% as memory pressure grows. For search workloads, \sys reduces latency by 10--16\% across different \texttt{nprobe} settings despite the random access patterns inherent in ANN search.

\begin{figure}
\centering
\includegraphics[width=\columnwidth]{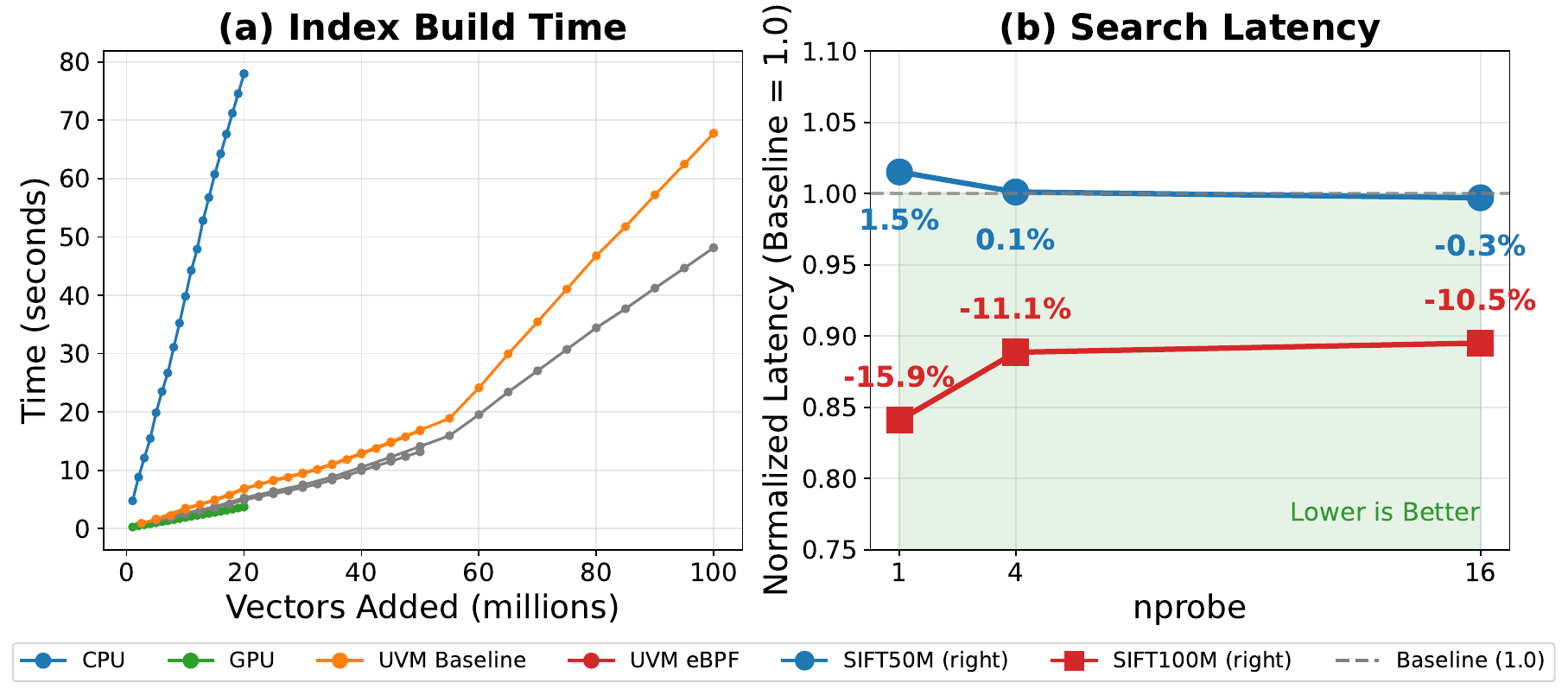}
\vspace{-1ex}
\caption{Faiss index build time and query throughput under oversubscription. \sys adaptive prefetch reduces build time by 21--29\% and query latency by 10--16\% compared to default UVM.}
\label{fig:faiss-perf}
\end{figure}

Across these applications, programmable region-based placement and prefetch policies in \sys exploit workload structure (experts, KV-cache, graph structure, index layout) that existing UVM policies and framework-level offload mechanisms cannot leverage.

\subsection{RQ2: Multi-Tenant Memory, Bandwidth, and Scheduling}

We evaluate \sys on multi-tenant GPU workloads, comparing per-tenant policies against global and framework-managed approaches.

\subsubsection{Microbenchmarks}

\begin{figure}
\centering
\includegraphics[width=\columnwidth]{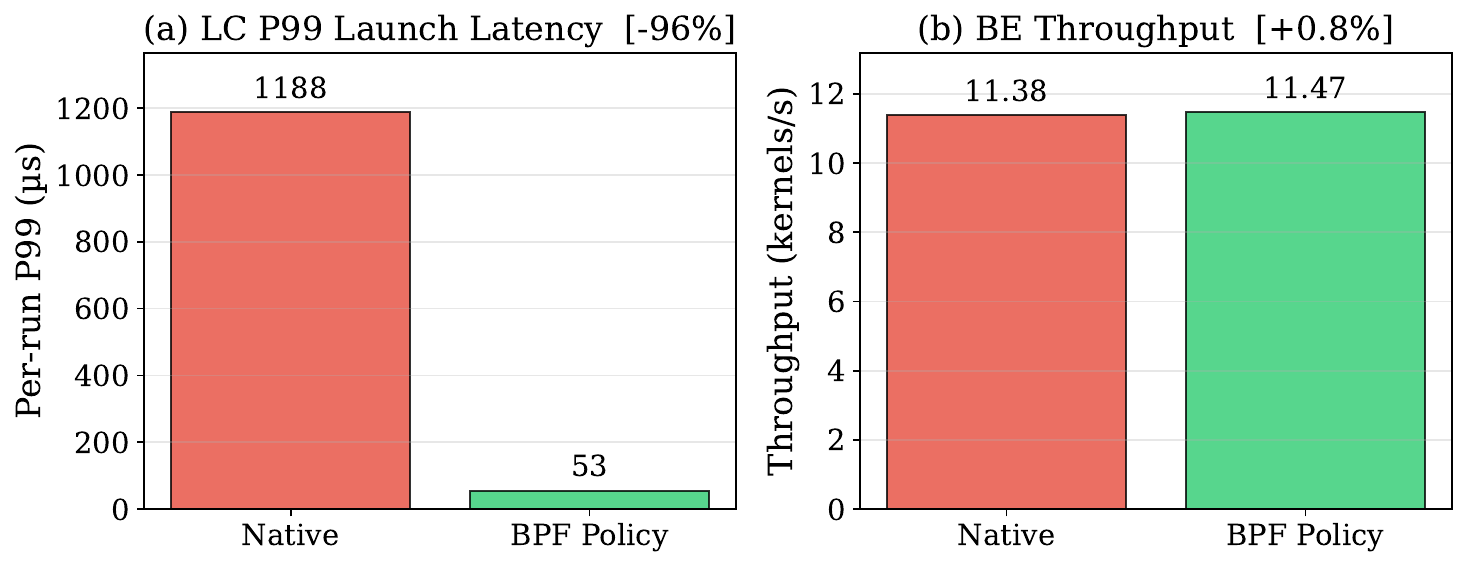}
\vspace{-1ex}
\caption{Multi-tenant GPU scheduling with \sys BPF struct\_ops policies on compute-bound workloads. \sys reduces latency-critical (LC) P99 launch latency by 95\% while maintaining best-effort (BE) throughput.}
\label{fig:scheduler-latency}
\end{figure}

\paragraph{Compute-bound Timeslice Scheduler (LC + BE).}
We evaluate \sys's BPF struct\_ops scheduling policies on \emph{compute-bound} multi-tenant workloads where the bottleneck is GPU SM time rather than memory bandwidth. 2 latency-critical (LC) processes and 4 best-effort (BE) processes run concurrently, each with 4 CUDA streams submitting 50 compute kernels. We compare the native scheduler against \sys with differentiated timeslices (LC: 1s, BE: 200$\mu$s) using a gpreempt-style preemption policy. \Cref{fig:scheduler-latency} shows that \sys reduces LC P99 launch latency by 95\%, demonstrating scheduling stability. BE throughput remains unchanged, confirming that \sys reduces LC tail latency without sacrificing BE throughput.

\paragraph{Memory-bound Memory Priority Differentiation.}
We evaluate \sys's memory policies for priority differentiation on \emph{memory-bound} workloads under UVM oversubscription using HotSpot~\cite{che2009rodinia} (spatial locality), GEMM from PolyBench~\cite{grauer2012auto} (compute-intensive), and K-Means~\cite{gu2020uvmbench} (sparse access). Each experiment runs two concurrent processes competing for GPU memory. Policy labels use priority values 0--100 (lower value = higher priority): \texttt{Prefetch(lo,hi)} prefetches pages for processes in that range, while \texttt{Evict(lo,hi)} selects eviction victims accordingly. We compare against: (1) no-policy execution, (2) single-process runtime (Single 1$\times$), and (3) theoretical optimum (2$\times$Single 1$\times$).

\begin{figure}
\centering
\includegraphics[width=\columnwidth]{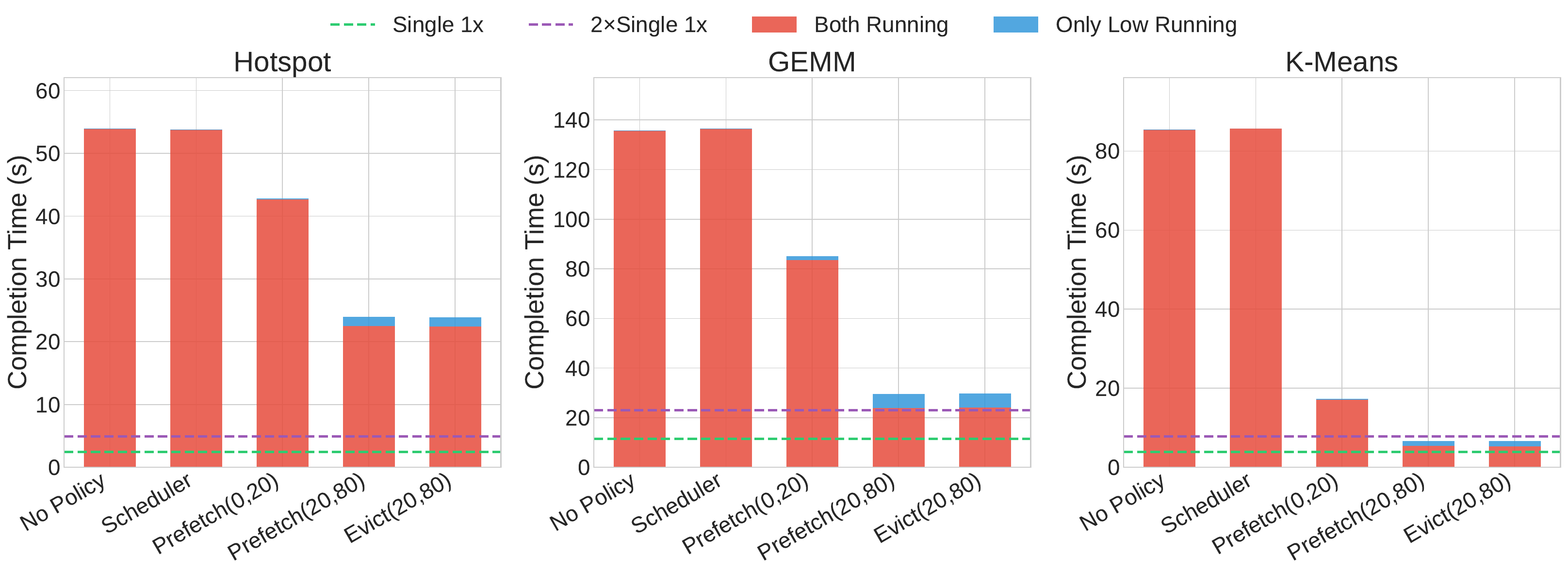}
\vspace{-1ex}
\caption{Multi-tenant memory priority differentiation on oversubscribed UVM workloads. \sys memory policies improve total completion time by 55--92\% and enforce priority differentiation; scheduler policies are ineffective on these memory-bound workloads. Single 1$\times$: ideal single-process time; 2$\times$Single 1$\times$: theoretical lower bound.}
\label{fig:all-kernels-priority}
\end{figure}

\Cref{fig:all-kernels-priority} shows that without policies, both processes degrade severely due to memory thrashing. With \sys memory priority policies, total completion time improves by 55--92\%, and high-priority processes complete 6--19\% faster. Critically, for these \emph{memory-bound} workloads, GPREEMPT-style~\cite{fan2025gpreempt} scheduler timeslice policies are \emph{ineffective} (<1\% improvement) because the bottleneck is UVM page faults and PCIe bandwidth, not GPU compute time. This contrasts with compute-bound workloads where scheduler policies are effective. \sys's layered runtime allows applying appropriate policies based on workload characteristics.

\subsubsection{Case Studies}

We apply the policy primitives from the microbenchmarks to realistic multi-tenant scenarios.

\paragraph{Two-Tenant: High-Priority LLM Inference + Background Training.}
We run two tenants sharing a single GPU: T1 is a high-priority \texttt{llama.cpp} inference service (LC workload) serving 100 ShareGPT prompts at 0.2 RPS, and T2 is a background PyTorch GNN training job (BE workload) with 8M nodes requiring 36GB peak memory (oversubscribed). We compare default UVM (driver FIFO/LRU) against \sys per-tenant memory policies where the LC tenant receives prefetch priority and the BE tenant yields bandwidth during inference.

\begin{figure}
\centering
\includegraphics[width=\columnwidth]{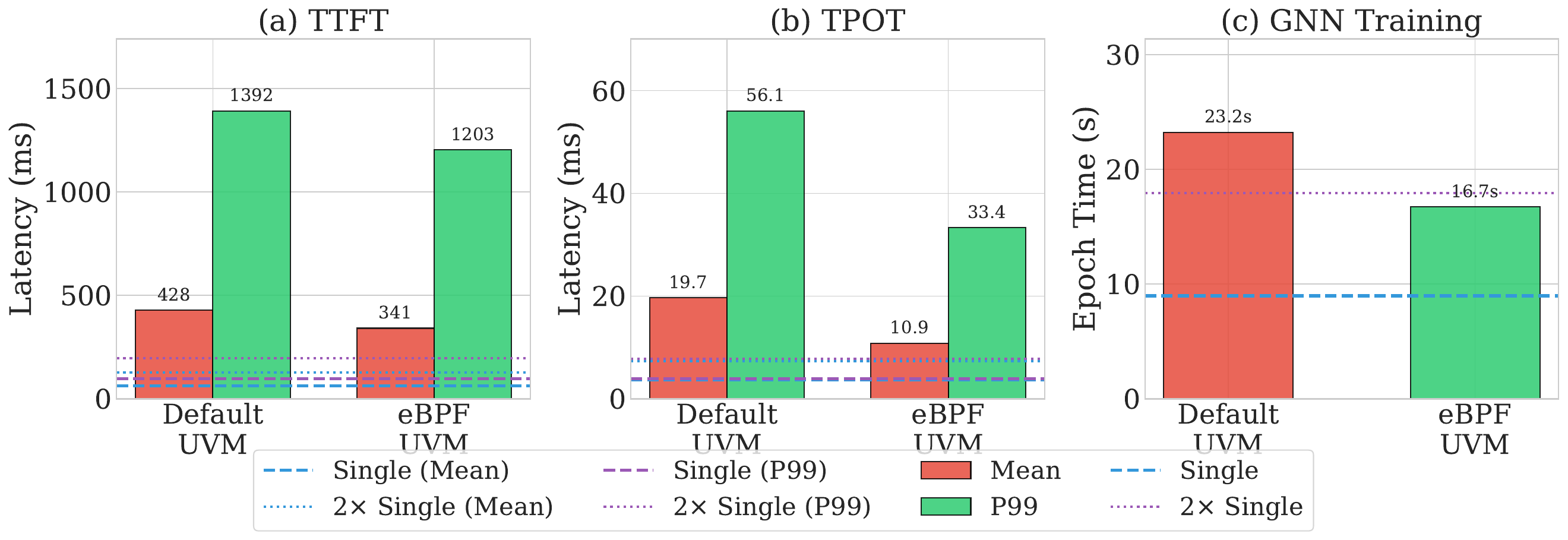}
\vspace{-1ex}
\caption{Two-tenant co-location: \texttt{llama.cpp} inference (LC) + GNN training (BE). \sys achieves mutual improvement: LC latency better than single-tenant performance while BE training improves by 28\%. Tenant-aware policies reduce contention rather than prioritizing one workload at the expense of another.}
\label{fig:two-tenant}
\vspace{-2ex}
\end{figure}

\Cref{fig:two-tenant} shows the results. \sys achieves mutual improvement for both tenants: LC latency reduces TPOT by 40--45\% and TTFT by 14--20\%, while GNN training simultaneously improves by 28\%, approaching the ideal fair share baseline. This shows \sys reduces contention for both workloads rather than favoring one over the other.

\sys's per-tenant policies improve both tail latency and fairness, benefiting high-priority tenants while reducing memory thrashing for background jobs.

\begin{figure*}[!t]
\centering
\includegraphics[width=\textwidth]{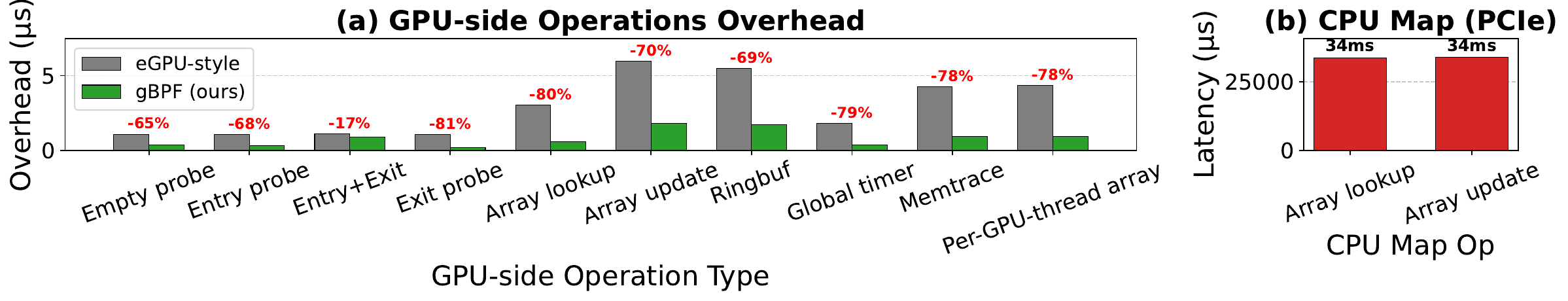}
\caption{Device-side microbenchmark overhead. (a) GPU-side operation overhead comparing eGPU-style naive injection vs \sys's SIMT-aware execution. (b) eGPU CPU map access latency via PCIe, motivating \sys's hierarchical map design.}
\label{fig:microbench}
\end{figure*}

\subsection{RQ3: Programmability and Mechanism Overhead}

We evaluate \sys's programmability by examining policy complexity and reuse, and quantify the efficiency of its core mechanisms and observability capabilities.

\paragraph{Policy Building Blocks.}
\Cref{tab:support-matrix} shows policy building blocks with lines of code and execution domain. These components can be composed to construct policies; for example, multi-tenant benchmark in RQ2 combines Quota LRU, Priority Tree Prefetch, and Dynamic Timeslice. All policies attach to unmodified frameworks via runtime instrumentation, requiring only tens to hundreds of lines.

\begin{table}[t]
\centering
\small

\begin{tabularx}{\columnwidth}{@{} X l c c @{}}
\toprule
\textbf{Policy} & \textbf{Tag} & \textbf{Domain} & \textbf{LOC} \\
\midrule
Global FIFO Eviction & Memory Eviction & Host & 145 \\
Global LFU Eviction & Memory Eviction & Host & 304 \\
Multi-tenant Quota LRU & Memory Multi-tenant & Host & 472 \\
Adaptive Seq. Prefetch & Memory Prefetch & Host & 375 \\
Stride Prefetch & Memory Prefetch & Host & 472 \\
GPU L2 Stride Prefetch & Memory Prefetch & Device & 45 \\
Tree-based Prefetch & Memory Multi-tenant & Host & 454 \\
Dynamic Timeslice & Scheduling & Host & 408 \\
Preemption Control & Scheduling & Host & 925 \\
MaxSteals (CLC) & Scheduling Block & Device & 16 \\
LatencyBudget (CLC) & Scheduling Block & Device & 19 \\
\bottomrule
\end{tabularx}
\caption{Policy support matrix. Domain indicates where the policy executes: Host (driver eBPF), Device (GPU eBPF), or both.}
\label{tab:support-matrix}
\end{table}

\subsubsection{Host Runtime Overhead}

We quantify the overhead of the \sys driver runtime independent of policy logic using GEMM from PolyBench~\cite{grauer2012auto} and HotSpot~\cite{che2009rodinia} (1.1$\times$ oversubscription, 35~GB working set) on RTX 5090. With hooks enabled but no policy attached, the overhead is less than 0.2\%. 

\subsubsection{Low-overhead GPU Observability}

We evaluate device-side observability tools built on \sys (\Cref{tab:obs-overhead}) on NVIDIA P40 GPU using llama.cpp prefill (Llama 1B, IO3, sequence-mixed). Tools include \texttt{kernelretsnoop} for per-thread block finish timestamp recording, \texttt{threadhist} for load imbalance detection, and \texttt{launchlate} for GPU kernel launch latency measurement. Overhead is measured as token/s degradation during prefill. These tools show significantly lower overhead than NVBit~\cite{villa2019nvbit} (3--14\% vs 85--87\%) due to \sys's warp-uniform execution avoiding divergent control flow and unnecessary memory accesses.

% \texttt{gpu\_sched\_trace}, \texttt{prefetch\_trace}, and \texttt{eviction\_trace} provide host-side observability for scheduling and memory management policies (\Cref{tab:obs-overhead}).
\begin{table}[t]
\centering
\small
\begin{tabularx}{\columnwidth}{@{} X l c c c @{}}
\toprule
\textbf{Tool} & \textbf{LOC} & \textbf{Domain} & \textbf{\sys} & \textbf{NVBit} \\
\midrule
kernelretsnoop & 153 & Device & 8\% & 85\% \\
threadhist & 89 & Device & 3\% & 87\% \\
launchlate & 347 & Host+Device & 14\% & 93\% \\
% mem\_trace & 110 & Device & 5\% & --- \\
% threadscheduling & 402 & Device & 7\% & --- \\
% gpu\_sched\_trace & 586 & Host (kprobe) & <1\% & --- \\
% prefetch\_trace & 430 & Host (kprobe) & <1\% & --- \\
% eviction\_trace & 398 & Host (kprobe) & <1\% & --- \\
\bottomrule
\end{tabularx}
\caption{\sys device-side observability tools and overhead comparison with NVBit.}
\label{tab:obs-overhead}
\end{table}

\paragraph{Device-side Runtime Overhead and Optimization.}
We evaluate device-side \sys mechanisms using a vector-add microbenchmark (32 elements, 10K iterations) from cuda-sample~\cite{nvidia_cuda_samples} on RTX 5090. \Cref{fig:microbench}(a) compares \sys's SIMT-aware warp-level execution against our prior eGPU baseline~\cite{egpu} that injects eBPF bytecode without SIMT-specific optimizations. The eGPU-style approach incurs overhead due to per-thread execution and uncoalesced memory access, while \sys's warp-uniform execution and coalesced map access reduce overhead by 60--80\% across operations. \Cref{fig:microbench}(b) shows CPU map access via PCIe is 6000$\times$ slower than GPU-side operations, motivating \sys's hierarchical map design that keeps hot state on GPU. We do not compare policy-level performance against eGPU or Neutrino because they are limited to read-only observability (\S\ref{sec:background}).

\section{Portability}
\label{sec:discussion}

\sys is designed for portability, though our current implementation targets NVIDIA GPUs. NVIDIA's production drivers bypass Linux's standard kernel interfaces, but our design aligns with generic Linux abstractions.
On the host side, for memory management we follow Linux's Heterogeneous Memory Management (HMM) and its \texttt{migrate\_vma} interface, already adopted by AMD ROCm with XNACK retryable faults. For scheduling, we abstract GPU queues based on Linux DRM scheduler's \texttt{drm\_sched\_entity}, mapping to NVIDIA TSGs and potentially AMD user-mode queues~\cite{rocm-gpu-memory,linux-gpusvm-doc,drm-gpu-sched,amdgpu-userq-doc}.
On the device side, our compiler generates vendor-neutral eBPF bytecode using SPIR-V, enabling future compatibility with AMD ROCm and Intel Level Zero. End-to-end portability requires vendor-specific runtime support and hardware tuning, which remain future work~\cite{spirv-ir-rfc,llvm-offload-design}.

\section{Conclusion}
\label{sec:conclusion}
We presented \sys, a cross-layer, verifier-backed policy runtime that treats the GPU driver and device as a single programmable OS subsystem. \sys exposes stable attach points for memory placement and scheduling, executes the same restricted eBPF IR on host and device, and shares state through cross-layer BPF maps, enabling low-overhead observability and adaptive policies. Across PyTorch, llama.cpp, vLLM, and Faiss, \sys-based policies transparently improve throughput by up to $4.8\times$ and reduce tail latency by up to $2\times$ relative to static heuristics, while instrumentation-only deployments incur low overhead. \sys demonstrates that kernel extensibility is an effective approach for GPU management.

% References: USENIX templates usually use plain or abbrvnat etc.
\bibliographystyle{plain}
\bibliography{cite}

\end{document}